\documentclass{elsart}
\usepackage{graphicx}

\usepackage{amssymb}

\begin{document}

\vskip 5 cm

\begin{frontmatter}



\title{Study of Neutron-Induced Ionization in \\ Helium and Argon Chamber 
  Gases}

 \author[ut]{D. Indurthy}
 \author[wisc] {A.R. Erwin}
 \author[fermi] {D.A. Harris}
 \author[ut]{S.E. Kopp\corauthref{myemail}} \hskip 1.0cm
 \corauth[myemail]{Corresponding author e-mail kopp@mail.hep.utexas.edu}
 \author[ut]{M. Proga}
 \author[ut] {R.M. Zwaska}
 \address[ut]{Dept. of Physics, University of Texas, Austin, Texas  
                                                       78712 U.S.A.}
 \address[fermi]{Fermi National Accelerator Laboratory, Batavia, Illinois 
                                                       60510 U.S.A.}
 \address[wisc]{Dept. of Physics, University of Wisconsin, Madison, Wisconsin 
                                                       53706 U.S.A.}

\vskip -.25 cm

\begin{abstract}

We present studies of Helium- and Argon-filled ionization chambers 
exposed to intense neutron fluxes from PuBe neutron sources 
($E_n=1-10$~MeV).  The sources emit about 10$^8$
neutrons per second.  The number of ion pairs in the
chamber gas volume per incident neutron is derived.  While limited in precision
because of a large gamma ray background from the PuBe sources, our results
are consistent with the expectation that the neutrons interact purely
elastically in the chamber gas.

\end{abstract}

\end{frontmatter}

\clearpage

\section{Introduction}

The Neutrinos at the Main Injector (NuMI) beamline at the Fermi
National Accelerator Laboratory \cite{numi} will generate an intense
$\nu_{\mu}$ beam from the decays of mesons produced in the 
collisions of 120~GeV protons on a graphite target.  The mesons 
are focused into a 675~m long evacuated 
volume where they decay to neutrinos.  The meson decays
$\pi/K \rightarrow \mu \nu_{\mu}$ produce an energetic muon for every
neutrino, allowing monitoring and validation of the neutrino beam 
focusing to be accomplished by monitoring of the muon flux.  As in 
several previous experiments, 
the muon flux and remnant hadron flux at the end of the decay volume
will be measured by arrays of ionization chambers
\cite{cern,blair,russians,k2k}.

\begin{figure}
\begin{center}
\includegraphics[scale=.75]{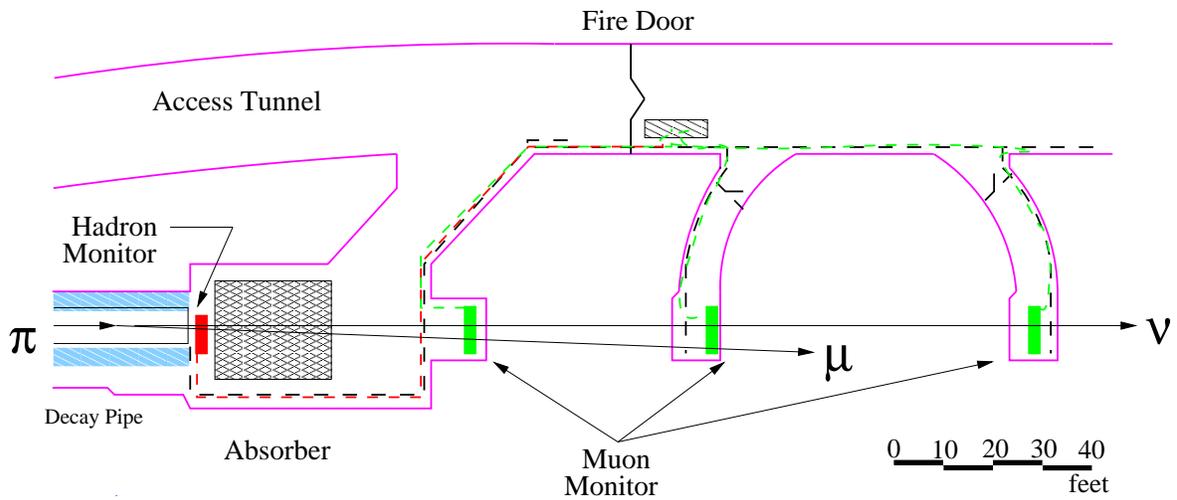}
\vskip -.2 cm
\caption{Location of the four ionization chamber arrays at the downstream 
end of the NuMI neutrino beamline.  Mesons decaying in the evacuated drift 
volume decay into muons and neutrinos.  Ion chamber arrays are located at four stations: 
upstream of the beam absorber and embedded in the earth in three ``alcoves.''
These arrays monitor the hadron and the muon components, respectively, of the beam.
\label{fig:alcove_layout}}
\end{center}
\end{figure}

In addition to charged particle fluxes, the ionization 
chambers are exposed to large fluences of neutrons, primarily 
produced in the aluminum/steel beam absorber upstream of the first
muon monitor (see Figure~\ref{fig:alcove_layout}).  The neutron
fluxes, as estimated by a {\tt GEANT}\cite{geant} simulation of
the beamline, are as great as 13$\times$10$^9$/cm$^2$/spill at the hadron 
monitor and 9$\times$10$^7$/cm$^2$/spill in the first muon station.
The neutron fluences are a factor of 10 larger than the 
charged particle levels the chambers should measure.  Most of the
neutrons are low energy, with a mean energy of 30~MeV.
The response of the ion chambers to a charged
particle flux has been studied with electron and
proton beams \cite{boosternote,atfnote}.  In this article, we study
the response of ion chambers to neutrons.

Previous neutrino beamlines have also utilized ionization chamber arrays to 
measure the muon beam profile and rate.
Because of background particles produced in the beam dump,
previous experiments have reported having to contend with
higher-than-anticipated fluxes \cite{baggett,nakaya}, or even to disregard the
rate information from the muon detectors altogether \cite{Auchincloss}.

Unlike charged particles, neutrons do not ionize the gas directly.  
Instead, fast neutrons interact with nuclei and create a charged recoil
nucleus.  This nucleus very quickly deposits
its energy through ionization.  The recoil particle can be produced
either in the chamber gas or in the plates of the chamber electrodes.

We studied the response of four
parallel plate ion chambers to neutron fluxes at the University
of Texas Nuclear Engineering Teaching Laboratory.  The neutrons
for this test are provided by PuBe radionuclide sources, which produce
1-10~MeV neutrons via the capture on Be of alpha particles emitted 
by the Pu \cite{Knoll}.  The total source activity is approximately 55~Ci, 
leading to approximately 1.5$\times$10$^8$~neutrons/sec.  The objective 
of this test is to determine the ionization due to neutron 
flux, an effective measurement of the cross section for neutron scattering
times the ionization left by the recoiling particles.

\section{Experimental Setup}
\label{Setup}

\begin{figure}
\begin{center}
\includegraphics[scale=1.25]{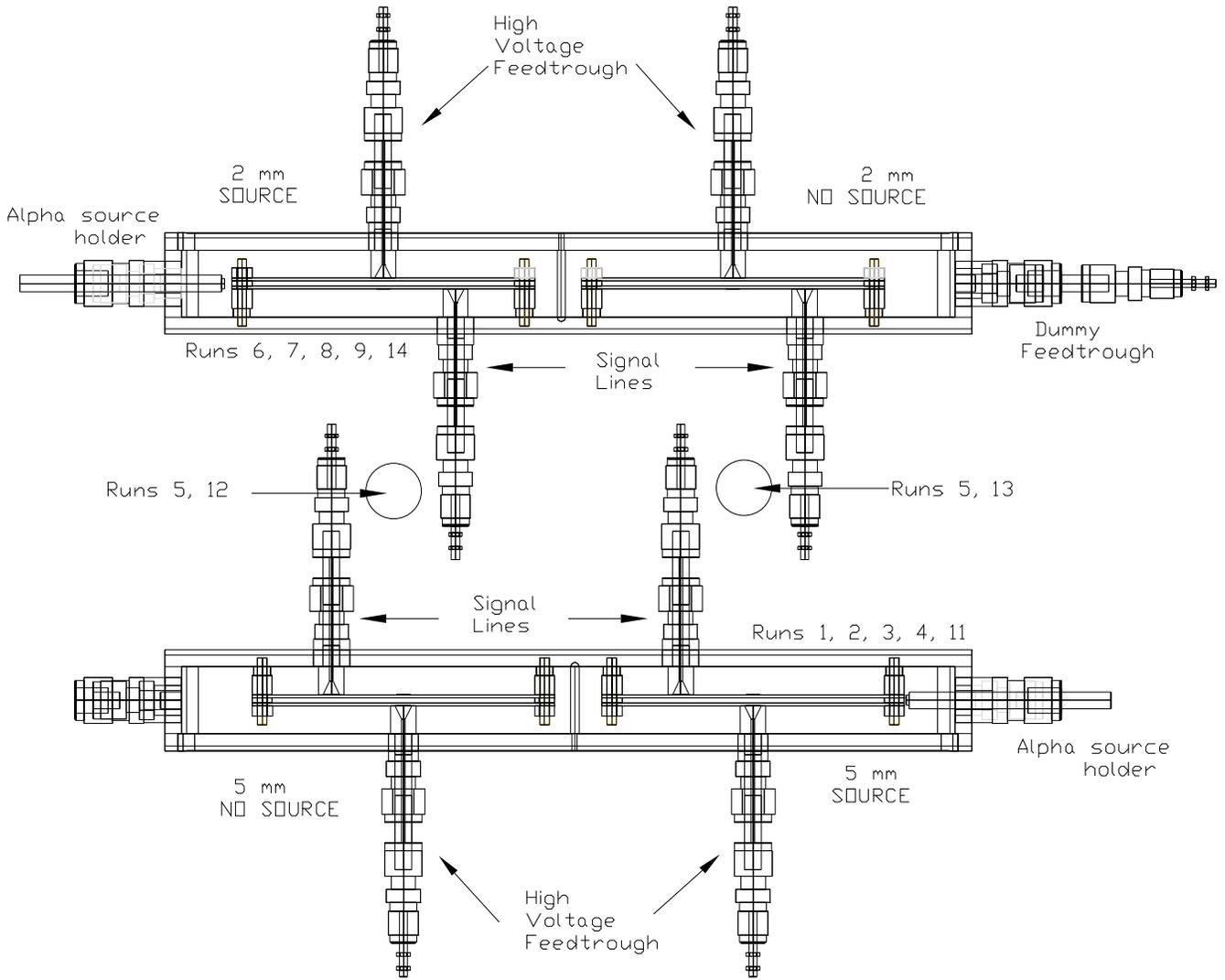}
\caption{The apparatus used in this test.  Two gas vessels, each
with two parallel plate ionization chambers, are placed approximately
20~cm apart.  One vessel has 2~mm gap chambers and one has 5~mm gap 
chambers.  The placements of the PuBe sources near the gas vessels
for the runs in Table~\ref{tab:runs} are indicated.
\label{fig:setup}}
\end{center}
\end{figure}

The four parallel plate ion chambers are housed in two 
separate Aluminum gas vessels (see Figure~\ref{fig:setup}).  
One vessel contains two chambers with 2~mm gas electrode separation, 
the other has two chambers with 5~mm electrode separation.  
Electrical feedthroughs to the ion chambers 
are made using PEEK insulating plastic and Aluminum compression fittings.  
A grounded Aluminum mesh inside the vessel separates
the two chambers.  The parallel plate ion chambers are fabricated from 
1~mm thick  100$\times$100~mm$^2$ ceramic plates with electrodes made 
of Pt-Ag alloy, 10$\mu$m in thickness.  The sense plane has a central
80$\times$80~mm$^2$ collection electrode with a 10~mm wide guard ring.
The chamber materials are identical to that to be used in the NuMI
beam monitoring chambers.

One ion chamber in each vessel has a 1~$\mu$Ci Am$^{241}$ $\alpha$ 
source mounted such that the $\alpha$ particles enter its gas volume.
The alpha sources permit calibration of temporal 
effects such as barometric pressure variations.
Each vessel also has a signal feedthrough with no internal connection
to an ion chamber.  These ``dummy'' feedthroughs serve to check
for spurious signals.  

Signals and high voltage are delivered to the setup through a 
shielding wall via coaxial cable with kapton insulator.  
The exterior of all feedthroughs is potted with epoxy, then 
wrapped in Aluminum foil to provide a complete grounded shield.
This exterior shielding reduces stray electric fields from the
high voltage feedthroughs to the signal feedthroughs which can 
drift ionized charges in the surrounding air to the signal lines.

The chambers are tested under Helium or Argon
gas flow at 1000~mbar.  The gas used in this test is rated as 99.998\% pure 
and is passed through an Oxygen getter\footnote{Matheson model MTRP-0012-XX} 
before going to the chambers which are connected
in series to the gas delivery.  The exhaust gas is monitored with 
an Oxygen analyzer\footnote{Illinois Instruments, model 2550}, showing 
0.5-1.5~ppm of O$_2$ impurity level during the tests.

The ionization current of each chamber is read by a 
picoammeter.\footnote{Keithley model 480}
The nominal sensitivity of the digital display of a 
picoammeter is 1 pA, however we found that by using the 
voltage output of the current-to-voltage amplifier we could achieve 
a sensitivity of better than 0.1 pA for steady currents.  The reason for
the 1~pA rating of the picoammeter appears to be drift in
the internal amplifier circuit which occurred sporadically during
the tests.  With frequent monitoring of the signals, we could discard 
data runs in which non-physical drifts are observed.  The picoammeter 
output is digitized by an ADC (least significant bit is equal to 
0.06~pA) and read into a computer.

\section{Neutron Source}
\label{Source}

The neutrons for this test are obtained from six individual PuBe 
sources.  Four of the sources are $^{239}$Pu totalling
8.99 Curies.  The two others provide 46~Ci of $^{238}$Pu.  
The neutron yield from the $^{239}$Pu is measured to be
57 per 10$^6$ alpha decays, but calculations suggest 
65 per 10$^6$ $\alpha$ decays \cite{Geiger}.  Given that 
yields may vary for specific alloy samples \cite{Knoll}, we take
$(60\pm5)\times10^{-6}$~neutrons/$\alpha$.  The theoretical 
neutron yield for $^{238}$Pu is $(79\pm5)\times10^{-6}~n/\alpha$ 
\cite{Anderson}.  We estimate all sources combined to emit 
$(1.57\pm0.13)\times10^{8}$ neutrons/sec.

During the test the neutron sources are moved to several locations
around the gas vessels.  
The neutron flux through each chamber in these runs is estimated 
using the source activities and the solid angle subtended by
the chambers.
In performing the solid angle integral over the
ion chamber faces, each PuBe source is assumed to be point-like.
The neutron flux is quoted in units of pA, where 
1pA=6.2$\times$10$^6$~neutrons/sec.  The fluxes through each chamber
for each run are tabulated in Table~\ref{tab:runs}.

In Table~\ref{tab:runs} is calculated as well a cosine-weighted flux. 
A neutron traversing the chamber at normal incidence receives a 
weight 1.0, while a neutron entering at 45$^\circ$ to the normal 
receives an extra weight of $\sqrt{2}$.  
This cosine factor weights for the pathlength of the neutron passing 
through the ion chamber gas gap at non-normal incidence, which increases
its probability of interaction.
The error in the fluxes due to the estimation method is about 10\%  
for the smaller fluxes and 20\% for the largest, which are very
sensitive to the source positions.  

The PuBe source also emits gamma rays, which form a background for
the measurements of this paper.  In approximately 57\% of 
$^9$Be$(\alpha,n)^{13}$C captures which lead to an emitted neutron from
the PuBe source, the $^{13}$C$^{*}$ excited state emits a 4.4~MeV 
gamma \cite{croft}.  These $\gamma$'s may Compton scatter in our chamber
materials, yielding additional ionization from the Compton electron.
This background is discussed further in Section~\ref{Neutint}.

\section{Plateau Measurements}
\label{Plateau}

In each of the runs, the bias voltage for each chamber is ramped from 
zero volts until gas amplification is observed.  The plateau ionization 
currents from Runs 0 and 10, in which no neutron sources are present, 
are subtracted from the plateau ionization currents observed in subsequent 
runs to determine the neutron-induced ionization in the gas.
As expected, only the chambers with an alpha source show an appreciable 
signal during Runs~0 and 10.  All chambers 
show an additional signal when significant neutron fluxes are present.

\begin{figure}
\includegraphics[scale=.4]{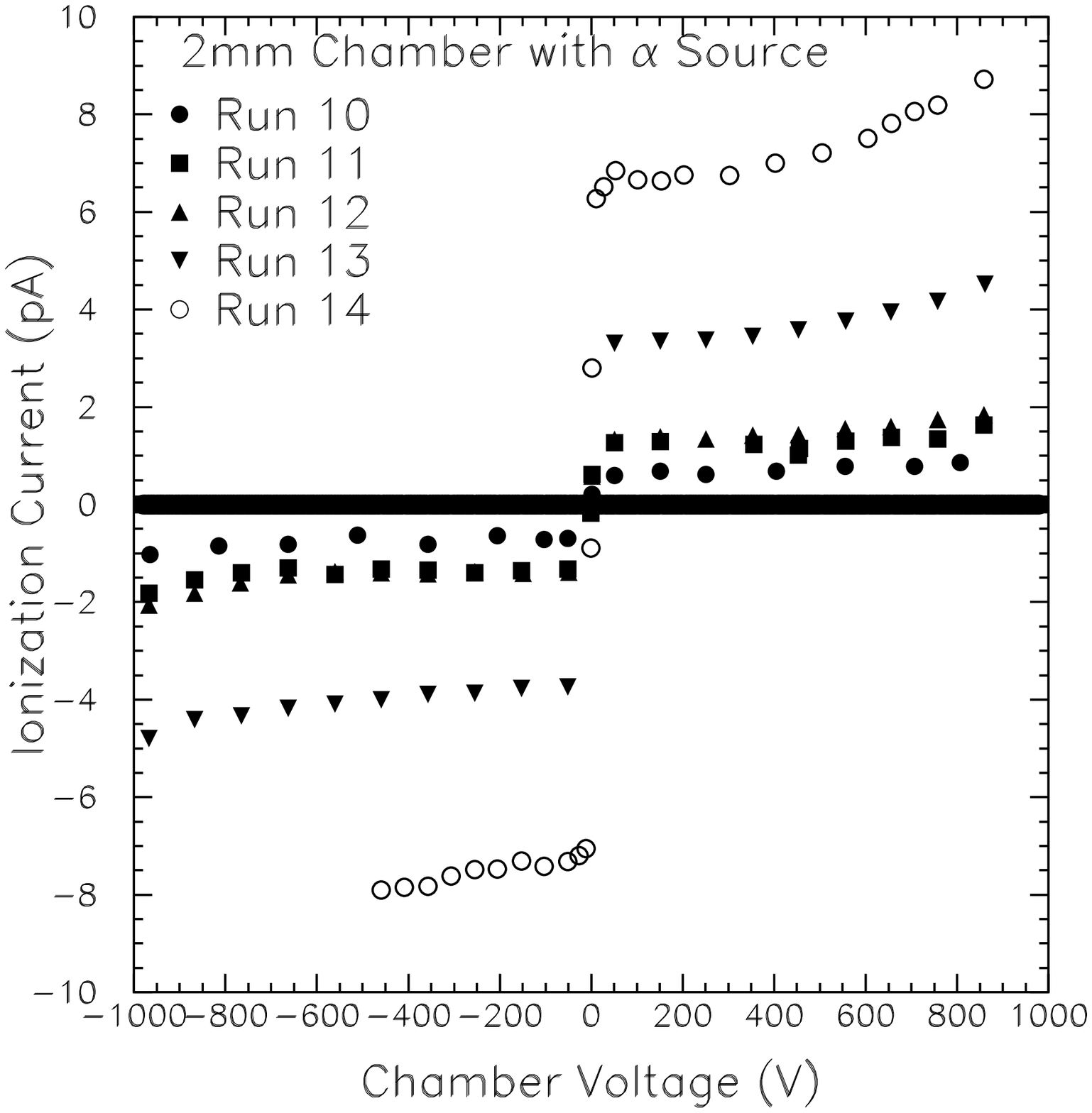}
\includegraphics[scale=.4]{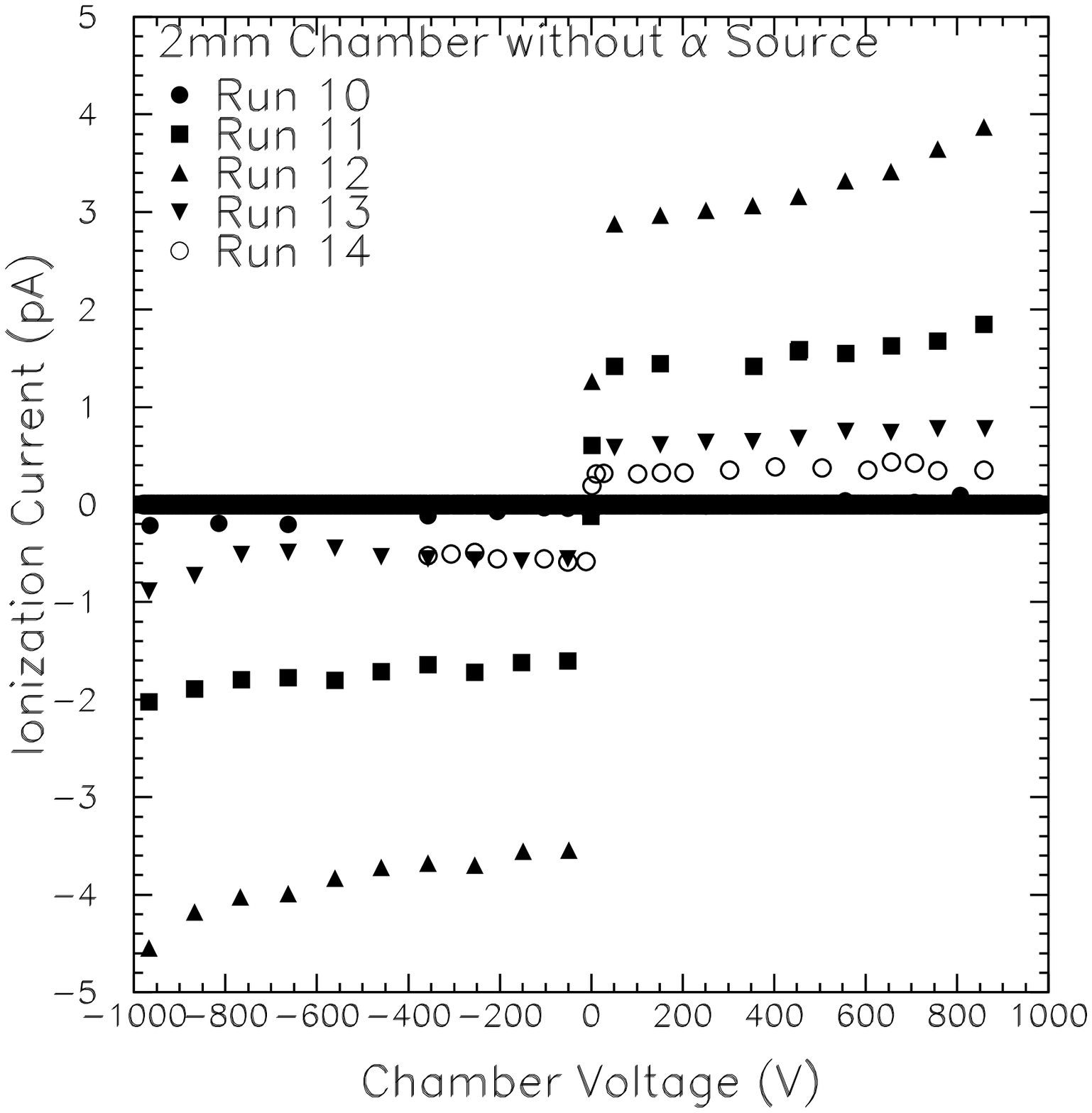}
\vskip -.5 cm
\includegraphics[scale=.4]{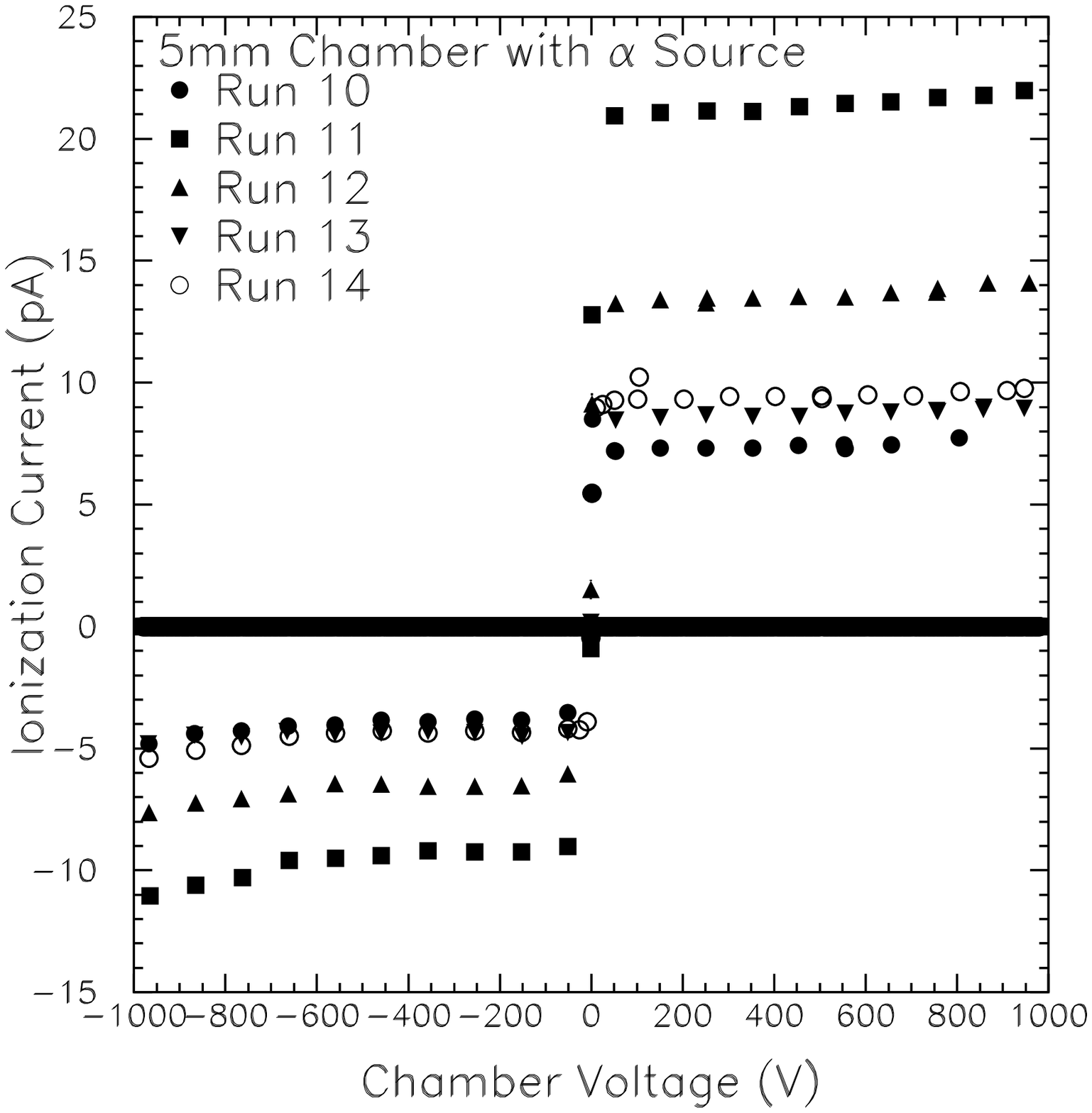}
\includegraphics[scale=.4]{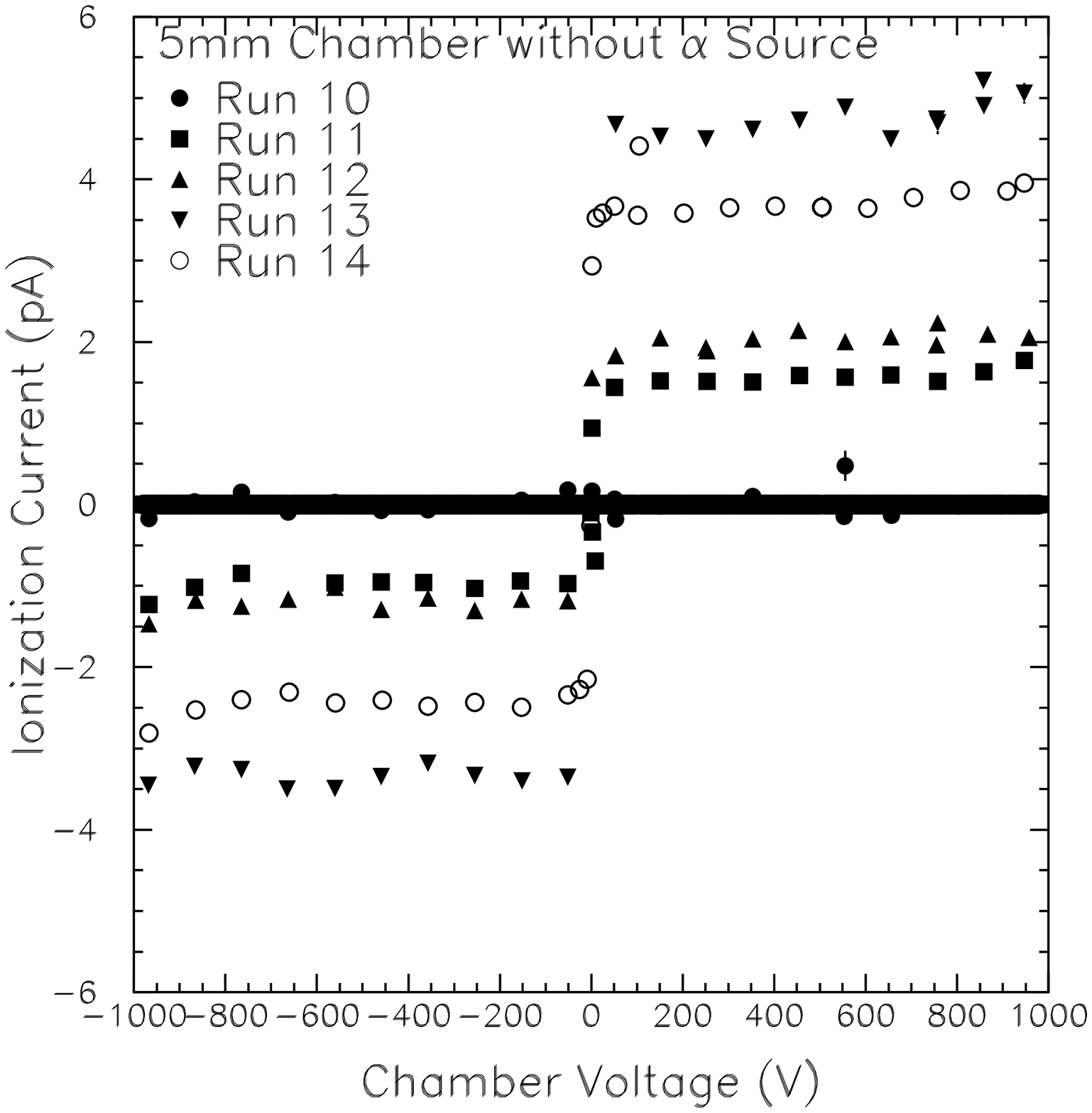}
\caption{Plateau curves for 2mm and 5mm chambers flushed with Argon gas for 4 
different neutron source placements.  
Each run corresponds to a neutron flux as indicated in Table~\ref{tab:runs}.
\label{fig:platsAr}}
\end{figure}

Because of voltage offsets in the the current-to-voltage amplifier in the
picoammeters, the definition of 'zero ionization current' is determined by 
measuring the chamber plateau curve with both positive and negative bias 
voltages and assuming the plateau values to be symmetrical about zero signal.  
This definition also calibrates away the effect of any contact potentials 
in the chambers, which cause ionization current to be collected even 
with zero applied bias voltage.  Figure~\ref{fig:platsAr}, for example,
shows the voltage plateau curves for the four chambers during different
runs taken with Argon gas flow.  While the 2~mm chambers are fairly symmetric
about zero ionization current, the 5~mm chambers show either some effect
of a contact potential or an offset in the picoammeters.

\begin{figure}
\includegraphics[scale=.4]{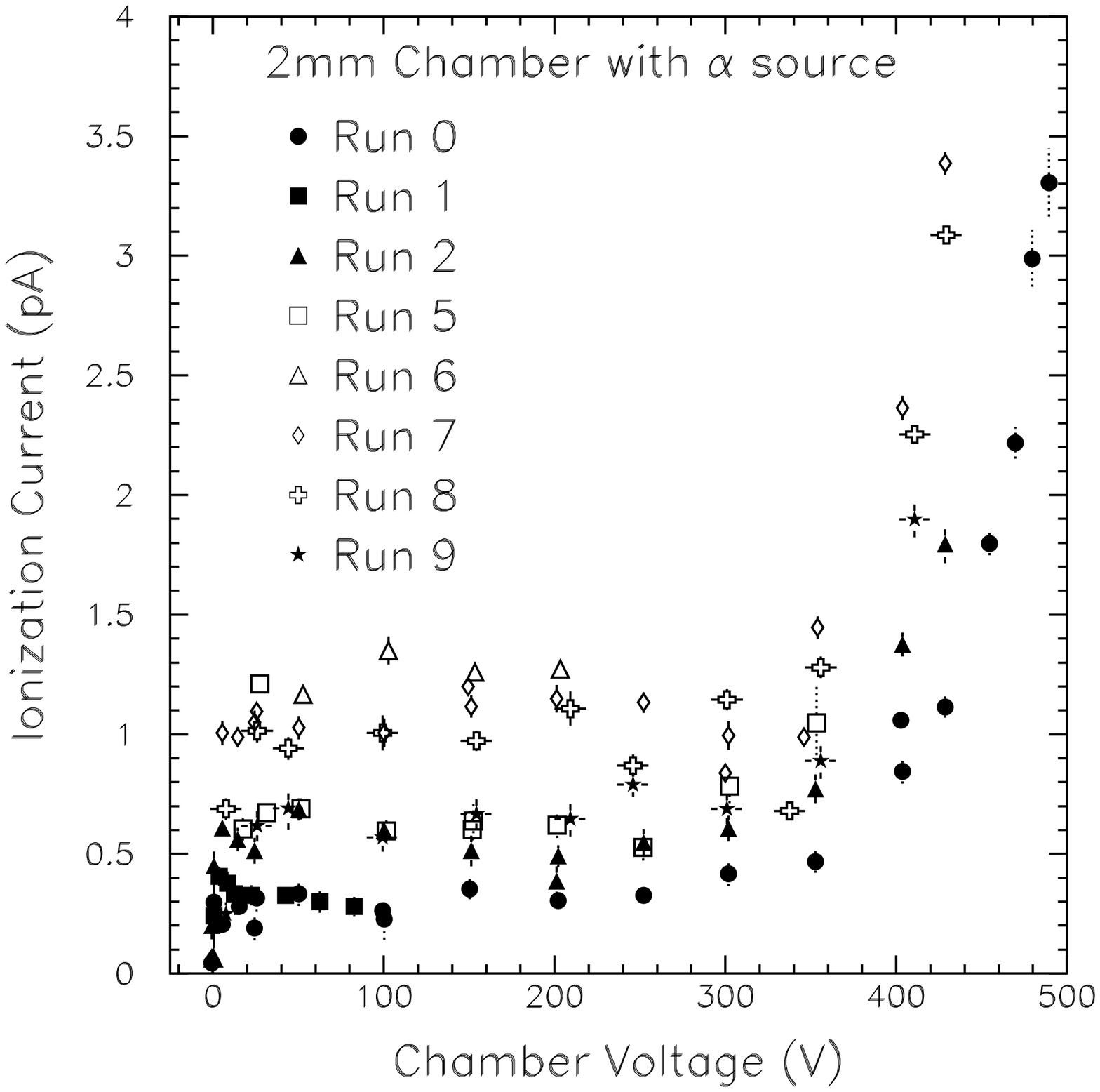}
\includegraphics[scale=.4]{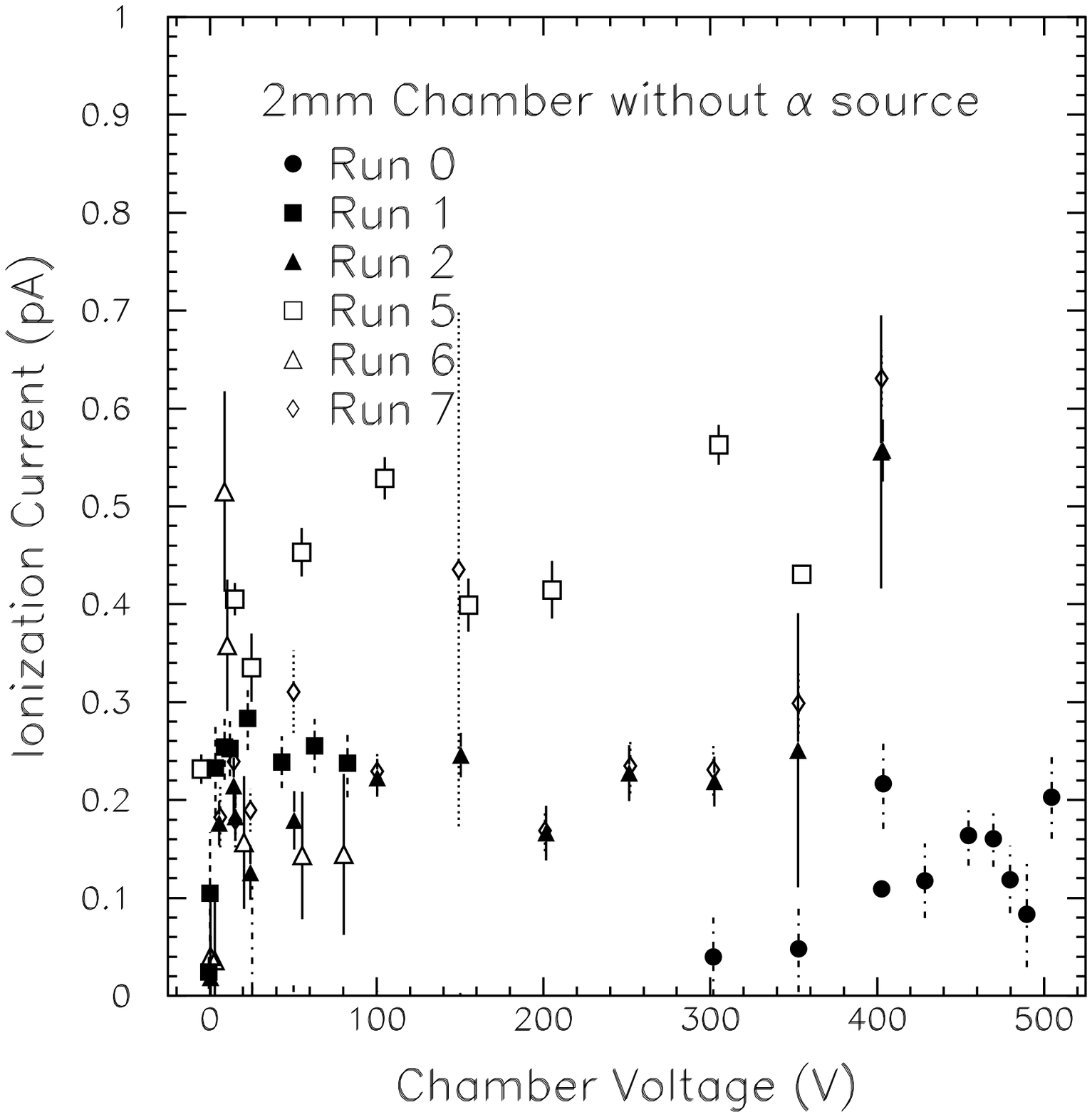}
\vskip -.5 cm
\includegraphics[scale=.4]{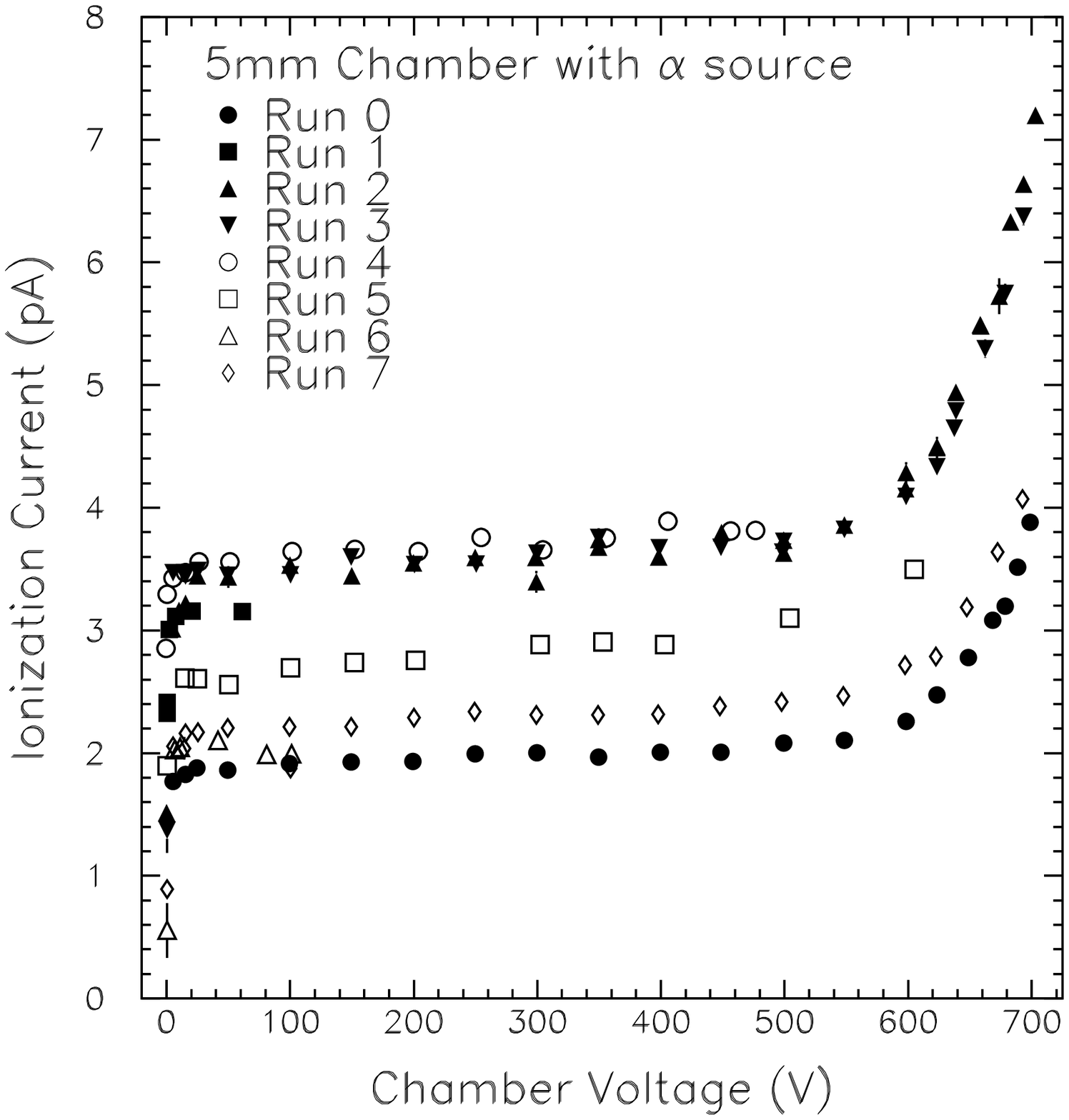}
\includegraphics[scale=.4]{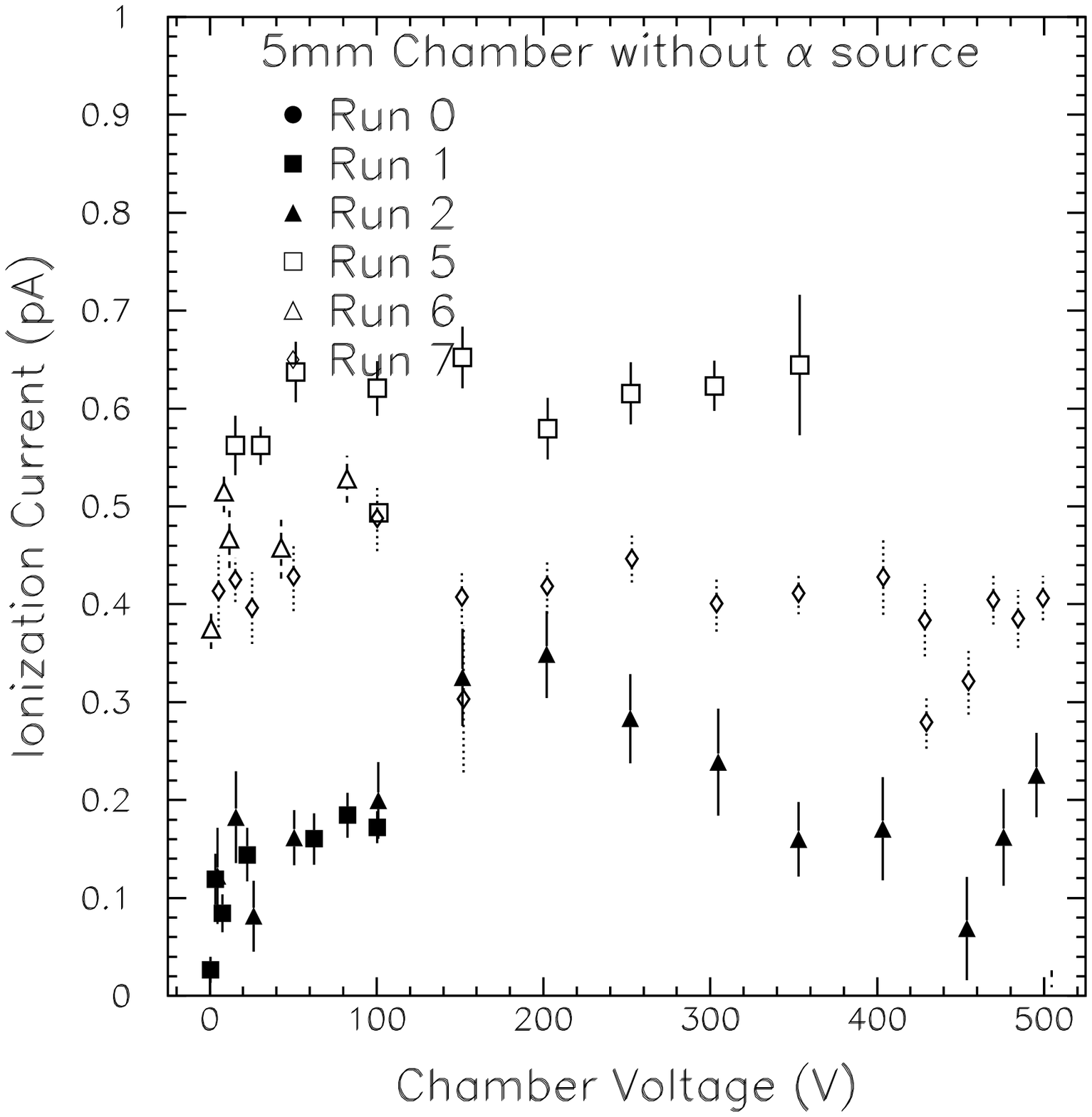}
\caption{Plateau curves for 2mm and 5mm chambers flushed with Helium gas 
for 9 different neutron source placements.  
Each run corresponds to a neutron flux as indicated in Table~\ref{tab:runs}.
\label{fig:plats}}
\end{figure}

Figures~\ref{fig:platsAr} and \ref{fig:plats} show the plateau curves for the 
runs taken with Argon and Helium gas flow, respectively.  In the Helium data,
the negative chamber bias data is not always available.  Therefore, only
the positive voltage portion of the He runs is plotted, where the voltage
plateaus have been scaled by the ratio of positive and negative ionization
currents observed in other runs.  The inavailability of the positive 
and negative polarities in some Helium runs leads to somewhat larger 
uncertainties in the ionization currents on plateau.

\begin{figure}
\begin{center}
\includegraphics[scale=.65]{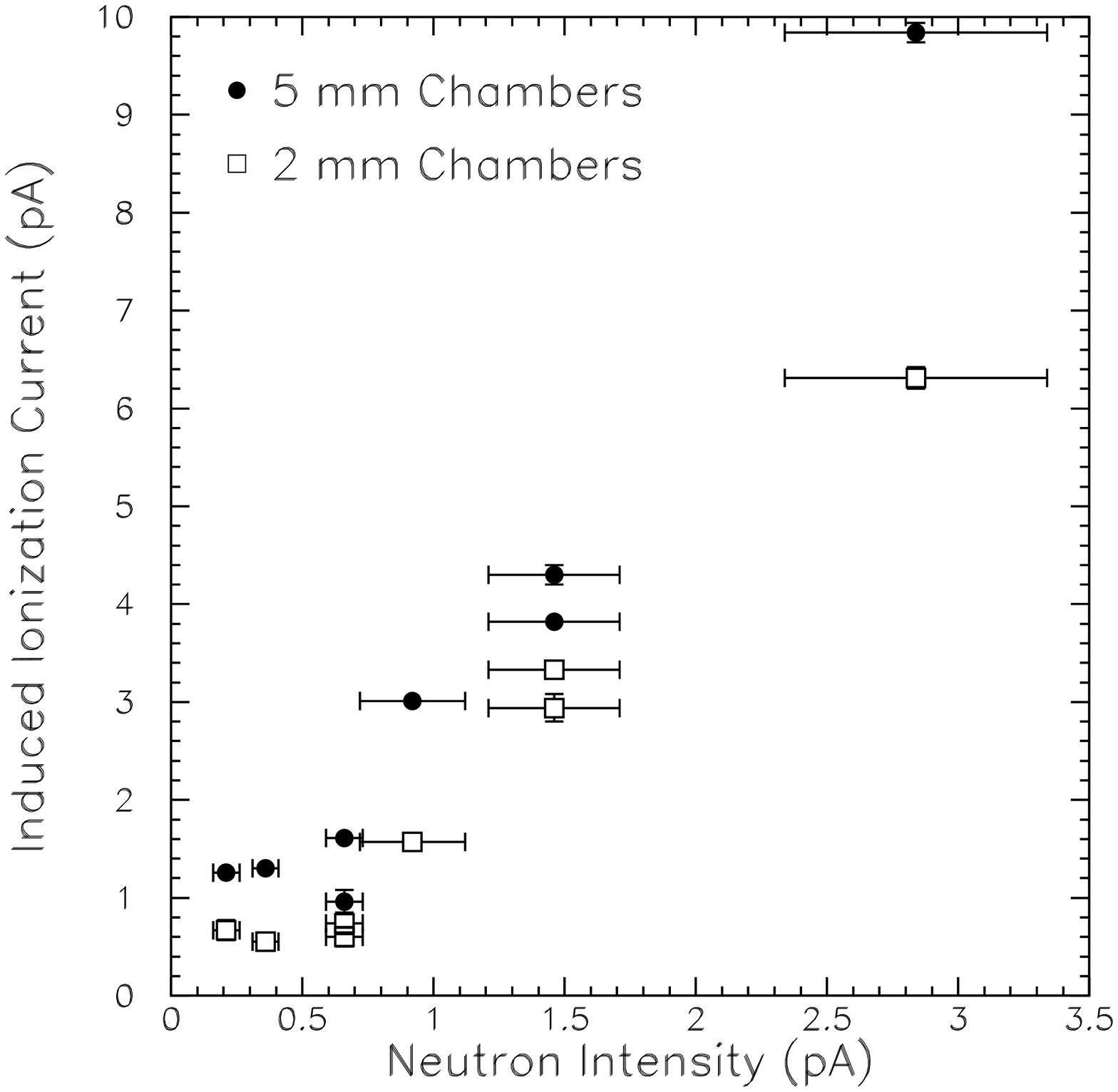}
\caption{Graph of the ionization current in Argon gas induced by the
PuBe source as a function of neutron intensity. The neutron intensity
is in units of 'pA', where 1~pA~=~6.2$\times$10$^6$~neutrons / sec. 
\label{fig:scatter4}}
\end{center}
\end{figure}

\begin{figure}
\begin{center}
\includegraphics[scale=.65]{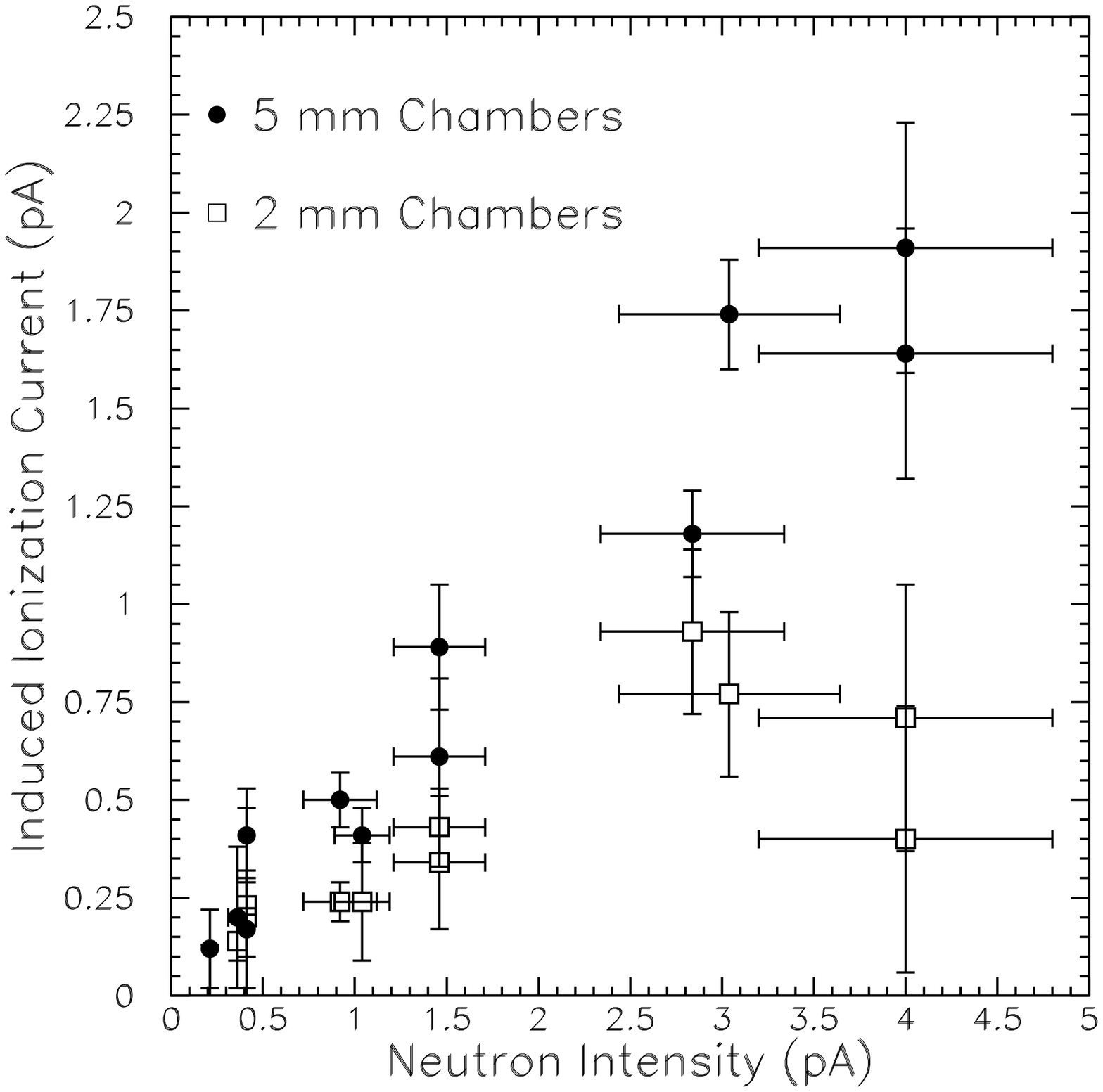}
\caption{Graph of the ionization current in Helium gas induced by the
PuBe source as a function of neutron intensity. The neutron intensity
is in units of 'pA', where 1~pA~=~6.2$\times$10$^6$~neutrons / sec. 
\label{fig:scatter2}}
\end{center}
\end{figure}

To deduce the ionization current due to the PuBe source, we subtract from the
ionization current of a given run the ionization current observed
in Run~0 (He data) or Run~10 (Ar data).  The chambers without alpha
sources are therefore more sensitive to small PuBe signals.  The 
ionization current attributed to PuBe-induced interactions is plotted as 
a function of neutron flux in Figures~\ref{fig:scatter4} and 
\ref{fig:scatter2} for the Argon and Helium data, respectively.

We have fit the data in Figures~\ref{fig:scatter4} and \ref{fig:scatter2}
to find the average PuBe source-induced ionization per neutron.  The fits
constrained to go through the origin gives slope values
indicated in Table~\ref{tab:results}.  Allowing the vertical 
intercept to float in the fit changes the slope in the Helium 
fits, for example, by only 0.03~ionizations/neutron, and the 
fitted vertical intercepts are consistent with zero 
(0.04$\pm$0.05 and 0.05$\pm$0.05 ionizations for the
5~mm and 2~mm chambers, respectively).

\section{Expected Rates}
\label{Neutint}

As stated above, some of the observed
ionization per neutron in Figures~\ref{fig:scatter4} and \ref{fig:scatter2}
and in Table~\ref{tab:results} is due to the accompanying 4.4~MeV $\gamma$'s 
from the PuBe source.  In this section we estimate the ionization rate 
induced by the PuBe source, accounting separately for (1) the ionization 
from recoil gas molecules scattered elastically by the PuBe neutrons; and
(2) the ionization from Compton electrons ejected by the 4.4~MeV 
$\gamma$'s.  ``Wall effects,'' whereby $\gamma$ rays are produced 
by inelastic interactions of neutrons in the chamber wall materials
\cite{schraube}, are not considered.  Similar estimates of neutron 
scattering in gaseous chambers have been used to estimate 
rates in drift chambers downstream of a beam dump \cite{E871}.  The
results of the calculations are summarized in Table~\ref{tab:expectations}.

\subsection{Neutron Rate}
 
The estimate of the ionization expected from neutron recoils in the 
gas first requires the measured total cross section 
for neutron scattering on Helium or Argon in the energy range 
2-40~MeV \cite{endf}, which are repeated in Table~\ref{tab:recoil}.  
These cross sections indicate an interaction 
length $\lambda=(1/n\sigma)$ of 1-2$\times10^4$~cm in the chamber gas.
The scattering is predominantly elastic below 10~MeV \cite{Knoll}.

The mean energy of an elastically-scattered nucleus is 
$\overline{E}_R = (2A/(1+A)^2)E_n$, where $E_R$ is the recoil nucleus 
energy, $E_n$ the incident neutron energy, and $A$ the 
target nucleus mass \cite{Knoll
}.  These recoil energies, 
along with the estimated ranges of these recoils in the chamber materials,
are listed in Table~\ref{tab:recoil}.  The ranges are extrapolated 
from measured ranges of alpha particles \cite{icru49}
using the scaling relation \cite{Evans} $R=R_{\alpha}(m/m_{\alpha})
(\sqrt{z_{\alpha}/z})$, where $m$ and $z$ are the mass and charge of the 
recoiling nucleus, $R_{\alpha}$ is the range of an alpha at
the same velocity as the nuclear recoil.  

The ranges of recoil Al, O, Pt, or Ag nuclei from the walls are 
insufficient to traverse the chamber wall materials and reach the 
chamber gas volume, with the exception of
Pt ions liberated very near the electrode surface.  Even the Pt ions
will contribute insignificant ionization in the chamber gas, 
since the Pt ions' energies are quite low.
In fact, the utility of Pt electrodes for reducing ``wall-effects'' 
in neutron interactions has been noted previously \cite{Coon}.

The ionization of the recoils created in the gas is then calculated.
Because of the short (1.2~mm) range of recoil Ar ions in Ar gas, we 
estimate their ionization as the recoil energy divided by
the energy $w=27$~eV to create an ionization in Ar gas \cite{icru},
yielding an average of 8900~ionizations.  For Helium recoils, 
we take $dE/dx=380$~keV/cm \cite{icru49} for $\overline{E}_R=$1.6~MeV 
alpha's, divided by $w=32$~eV to create an ion pair 
\cite{christophorou}, yielding 11,900~ionizations for a chamber with 1~cm
electrode spacing.

To estimate the average ionization per neutron, we multiply the 
probability for 
neutron scattering in the gas by the ionization deposited per 
recoil.  The results are given in Table~\ref{tab:expectations}.

\subsection{Gamma Rate}

To estimate the ionization caused by $\gamma$ rays emanating from the PuBe
source, we performed a parametric Monte Carlo calculation of $\gamma$'s Compton
scattering in the chamber wall materials, followed by energy loss and 
multiple scattering by the Compton-scattered electrons.  The ionization
induced in the chamber gas volume is calculated for those electrons which
arrive there.

Compton scattering of the 4.4~MeV $\gamma$'s 
occurs with approximately 1~barn cross section \cite{PDG}.  This yields
a probability of scattering in the Aluminum chamber walls of 25\% and
a probability of scattering in the ceramic plates of 11\%.  Including the
fact that only 57\% of neutrons are accompanied by a 4.4~MeV $\gamma$, this
yields 19\% of neutron events producing a Compton scatter in the chamber walls.

After the Compton scatter, the electron energy is calculated.  Its energy
loss is calculated using tabulated stopping power $dE/dx$ data \cite{icru37}.
Multiple scattering of the electron in the Aluminum wall and in the ceramic
is estimated as in \cite{PDG}.  
If the path through each material (aluminum or ceramic) is greater than the 
electron's range, the electron does not reach the chamber gas volume.  
Approximately 74\% of the Compton electrons are either ranged out in the 
chamber walls or are multiple scattered away from the gas volume.

The $dE/dx$ of the electrons arriving in the chamber gas volume is 
multiplied by the path length of the electrons through the gas volume.  
The electron pathlength through the gas can be larger than the electrode
separation due to their large multiple scattering.
The range of the electrons is $10^3-10^4$~cm
in He \cite{icru37}, so no electrons range out in the gas.  For the He gas, 
we find a mean of 3.7 (8.8) ion pairs created in the 2~mm (5~mm)
chamber.  
The results after multiplying by the probability 0.18$\times$0.26
to have a Compton electron, are shown in Table~\ref{tab:expectations}.

\section{Conclusions}

We have studied the signals induced in ionization chambers exposed to low 
energy (1-10~MeV) neutrons in the form of a PuBe source.  An ionization 
current from the source is observed whose magnitude is
(1.1$\pm$0.2$\pm$0.1)~ionizations/neutron/cm in the Helium gas, and 
(9.6$\pm$2.4$\pm$1.0)~ionizations/neutron/cm in the 
Argon gas, based on our observations at several neutron fluences and 
two chamber gas gaps.  The first uncertainty is that due to the estimate 
of the neutron fluxes, while the last uncertainty is due to the PuBe 
source activity into neutrons. 

Our measurements of the PuBe source-induced ionization are limited in 
precision because of the large number of $\gamma$ rays from the 
PuBe sources.  However, the neutron-specific signal inferred is
consistent with that expected from pure elastic scattering
of the neutrons in the chamber gas:  0.35~ion pairs per neutron per 
centimeter in He, and 0.75~ion pairs per neutron per centimeter in 
Argon.  Because of the problem of backgrounds from
the 4.4~MeV $\gamma$'s from the PuBe source, a study of the individual
pulse heights for each particle interaction in our chambers is desirable; 
such will be the subject of further study.

Our measurements indicate the NuMI beam monitors will have a background 
signal of order 20-30\% from neutron interactions in the chambers.
While the yield of ion pairs per neutron is only 2-3\% for that of
charged particles (assuming elastic scattering),
{\tt GEANT}\cite{geant}-based Monte Carlo estimates indicate that 
the hadron monitor and the first muon monitor 
will be exposed to the order of 10 times as many neutrons as 
signal particles (protons in the case of the Hadron monitor and 
muons in the case of the muon monitors).  
The neutron background signal will be 
much less correllated spatially with beam misalignments than the signal, 
so its effect will be important to model.

\section{Acknowledgements}

We thank the staff of the 
University of Texas Nuclear Engineering Department for providing the 
neutron source facility, and especially Sean O'Kelly and Bill Kitchen,
for their valuable assistance.  We thank Evan Fuller and James Hall for
assistance with the ionization chamber readout. 
This work was supported by the U.S. Department of Energy, DE-AC02-76CH3000
DE-FG03-93ER40757, and DE-FG02-05ER40896, and the Fondren Family Foundation.

\clearpage

\begin{table}
\begin{center}
\begin{tabular}{|c|c|c|c|c|c|c|}
\hline
      &  PuBe      &              & \multicolumn{4}{|c|}{ } \\
Run&Activity&Chamber&\multicolumn{4}{|c|}{Neutron Fluxes (pA) at Chambers} \\
 No.& (Ci)  & Gas&2mm $\alpha$&2mm no-$\alpha$&5mm $\alpha$&5mm no-$\alpha$ \\
\hline
0,10 & 0  & He,Ar & 0 & 0 & 0 & 0\\
1,11 & 46 & He,Ar & 0.10(0.21) & 0.83(0.92) & 2.23(2.84) & 0.15(0.36) \\
2 & 55 & He & 0.28(0.41) & 0.94(1.04) & 2.53(3.04) & 0.17(0.41)\\
3 & 55 & He & --         &  --        & 3.12(4.00) &  --       \\
4 & 55 & He & --         &  --        & 3.12(4.00) &  --       \\
5a,12 & 46 & He,Ar & 0.28(0.66) & 1.26(1.46) & 1.26(1.46) & 0.28(0.66)\\
5b,13 & 46 & He,Ar & 1.26(1.46) & 0.28(0.66) & 0.28(0.66) & 1.26(1.46)\\
6,12 & 46 & He & 2.23(2.84) & 0.15(0.36) & 0.10(0.21) & 0.83(0.92)\\
7 & 55 & He & 2.53(3.04) & 0.17(0.41) & 0.28(0.41) & 0.94(1.04)\\
8 & 55 & He & 3.12(4.00) & --         & --         & -- \\
9 & 55 & He & 3.12(4.00) & --         & --         & -- \\
\hline
\end{tabular}
\end{center}
\caption{Runs taken during the study of neutron interactions.  
The gas in the chambers is noted, as well as the expected 
neutron flux on each of the chambers during each run.  The neutron 
fluxes are quoted in units of 'picoamperes', where 
1~pA~=~6.2$\times$10$^6$~neutrons/sec.  Both the unweighted and 
cosine-weighted fluxes (in parentheses) are listed.
\label{tab:runs}}
\end{table}

\clearpage

\begin{table}
\begin{center}
\begin{tabular}{|c|c|c|}
\hline
Chamber		& \multicolumn{2}{|c|}{Observed Ion Pairs per Neutron} \\
Gap	& He Gas	  &	Ar Gas 	   \\ \hline\hline
2~mm		& 0.26 $\pm$ 0.07 & 2.4 $\pm$ 0.3  \\
5~mm		& 0.49 $\pm$ 0.06 & 3.7 $\pm$ 0.4  \\ \hline
1~cm$^a$	& 1.08~$\pm$~0.15 & 9.6~$\pm$~2.4  \\ \hline
\end{tabular}

$^a$Inferred from 2, 5~mm gap chambers.
\end{center}
\caption{Ionization current per neutron obtained from fits of the 
data in Figures~\ref{fig:scatter4} and \ref{fig:scatter2}.  The fitted
slopes are equivalent to ion pairs per neutron.
\label{tab:results}}
\end{table}

\begin{table}
\begin{center}
\begin{tabular}{|c|c|c|c|c|c|c|c|}
\hline
     &  &   & $\sigma_{tot}$ (b)\cite{endf} &  & \multicolumn{3}{|c|}{Particle Range in Media}\\
Element&$A$&$Z$& $E_n=2,5,10$ MeV & $\overline{E_R}/E_n $ & He (mm)& Ar (mm)& Pt ($\mu$m)\\
\hline
\hline
He 	  & 4  & 2  & 4.1, 2.3, 1.5  & 0.320 & 47   & --   & --  \\
Al  & 27 & 12  & 3.2, 2.3, 1.7 & 0.069 & 12.3 & 1.9  & 0.3 \\
Ar     & 40 & 18 & 4.3, 3.4, 2.2  & 0.048 & --   & 1.2  & -- \\
Ag    & 108& 47  & 6.5, 4.1, 4.4 & 0.018 & 2.6  & 0.27 & 0.1 \\
Pt  & 196& 78  & 5.8, 6.6, 5.2 & 0.010 & 0.74 & 0.13 & 0.1 \\\hline
\end{tabular}
\end{center}
\caption{Elastic scattering data for materials found in our ionization
chambers,   The total cross sections are given for the range 
2-10~MeV.  The average recoil energy is provided as a portion of the
neutron's energy.  Range is calculated for the typical fully-ionized 
recoil nucleus of a 5 MeV neutron in both gaseous Helium and Argon 
and in Platinum using data from \cite{icru49} and scaling laws 
from \cite{Evans}.
\label{tab:recoil}}
\end{table}

\begin{table}
\begin{center}
\begin{tabular}{|c|c|c|c|c|c|c|}
\hline
Chamber  & \multicolumn{6}{|c|}{Expected Ion Pairs per Neutron} \\
Gap	 & \multicolumn{3}{|c|}{He Gas}&\multicolumn{3}{|c|}{Ar Gas}  \\ \hline
   & $n$'s & $\gamma$'s & Total   &  $n$'s & $\gamma$'s & Total\\ \hline\hline
2~mm		& 0.07 & 0.18 & 0.25 & 0.15 & 1.8 & 2.0 \\
5~mm		& 0.18 & 0.44 & 0.62 & 0.37 & 4.4 & 4.8 \\
1~cm		& 0.35 & 0.84 & 1.19 & 0.75  & 8.4 & 9.2\\\hline
\end{tabular}
\end{center}
\caption{Expected ionization current per neutron decay from the PuBe source,
as described in Section~\ref{Neutint}. The estimates include the ion pairs
produced by elastic scatters of neutrons in the chamber gas ($n$'s), as well
as those arising from Compton scatters of subsequent cascade $\gamma$'s from 
the PuBe source ($\gamma$'s).  The total expected signals may be compared 
to the observed values listed in Table~\ref{tab:results}.
\label{tab:expectations}}
\end{table}

\clearpage

\clearpage

\clearpage

\clearpage

\clearpage

\clearpage 

\end{document}